\documentclass[aps,prl,twocolumn,superscriptaddress]{revtex4-1}
\usepackage{amsmath,amssymb,amsbsy,graphicx,xcolor,times,marvosym,color,bm,multirow,epstopdf}
\usepackage[latin1]{inputenc}
\usepackage{hyperref}
\usepackage{array}
\usepackage{footnote}
\newcommand*\diff{\mathop{}\!\mathrm{d}}
\newcommand{\mc}[1]{\mathcal{#1}}
\newcommand{\non}{\nonumber}
\newcommand{\la}{\langle}
\newcommand{\ra}{\rangle}
\newcommand{\cd}{c^{\dagger}}
\newcommand{\dg}{\dagger}

\begin{document}

\title{Quantization of the thermal Hall conductivity at small Hall angles}

\author{Mengxing Ye}
\affiliation{School of Physics and Astronomy, University of Minnesota,
  Minneapolis, Minnesota 55455, USA}
\affiliation{Kavli Institute for Theoretical Physics, University of
  California, Santa
Barbara, CA 93106, USA}

\author{G\'abor B. Hal\'asz}
\affiliation{Kavli Institute for Theoretical Physics, University of
  California, Santa
Barbara, CA 93106, USA}

\author{Lucile Savary}
\affiliation{Universit\'e de Lyon, \'{E}cole Normale Sup\'{e}rieure de Lyon, Universit\'e Claude Bernard Lyon I, CNRS, Laboratoire de physique, 46, all\'{e}e d'Italie, 69007 Lyon, France}

\author{Leon Balents}
\affiliation{Kavli Institute for Theoretical Physics, University of
  California, Santa
Barbara, CA 93106, USA}


\begin{abstract}
  We consider the effect of coupling between phonons and a chiral
  Majorana edge in a gapped chiral spin liquid with Ising anyons
  (e.g., Kitaev's non-Abelian spin liquid on the honeycomb lattice).
  This is especially important in the regime in which the longitudinal
  bulk heat conductivity $\kappa_{xx}$ due to phonons is much larger
  than the expected quantized thermal Hall conductance
  $\kappa_{xy}^{\rm q}=\frac{\pi T}{12} \frac{k_B^2}{\hbar}$ of the ideal isolated edge
  mode, so that the thermal Hall angle, i.e., the angle between the
  thermal current and the temperature gradient, is small.  By modeling
  the interaction between a Majorana edge and bulk phonons, we show
  that the exchange of energy between the two subsystems leads to a
  transverse component of the bulk current and thereby an {\em
    effective} Hall conductivity.  Remarkably, the latter is equal to
  the quantized value when the edge and bulk can thermalize, which
  occurs for a Hall bar of length $L \gg \ell$, where $\ell$ is a
  thermalization length. We obtain $\ell \sim T^{-5}$ for a
  model of the Majorana-phonon coupling.  We also find that the
  quality of the quantization depends on the means of measuring the
  temperature and, surprisingly, a more robust quantization is
  obtained when the lattice, not the spin, temperature is measured.
  We present general hydrodynamic equations for the system, detailed
  results for the temperature and current profiles, and an estimate
  for the coupling strength and its temperature dependence based on a
  microscopic model Hamiltonian.  Our results may explain recent
  experiments observing a quantized thermal Hall conductivity in the
  regime of small Hall angle, $\kappa_{xy}/\kappa_{xx} \sim 10^{-3}$,
  in $\alpha$-RuCl$_3$.
\end{abstract}

\date{\today}


\maketitle


Non-Abelian statistics is a deep generalization of quantum
statistics in two dimensions, in which the final state depends upon
the order in which exchanges of particles -- non-Abelian anyons --
are performed
\cite{moore1991nonabelions,read2000paired,wilczek2009}.  In addition
to its fundamental interest, this provides a powerful paradigm for
quantum computing, allowing for fault-tolerant processes
\cite{kitaev2003fault,nayak2008non}. The main platforms in which
non-Abelian topological phases have been sought are the $\nu=5/2$
Fractional Quantum Hall Effect (FQHE)
\cite{moore1991nonabelions,read2000paired}, where non-Abelian anyons
are suspected but have not been established, 
and hybrid
semiconductor-superconductor structures, to which quantum computing
groups are devoting massive efforts \cite{lutchyn2018majorana}, but
where confirmation is still awaited.

A third possible route to non-Abelian anyons is via a quantum spin
liquid \cite{savary2016quantum}.  In his seminal work
\cite{kitaev2006anyons}, Kitaev presented a spin-1/2 model on the
honeycomb lattice with bond-dependent anisotropy which, in a
magnetic field, realizes a non-Abelian topological phase.  This
phase hosts {\em Ising anyons}, topologically the same anyon type
which is targeted by the hybrid efforts.  A key and general
characteristic of a topological phase is the {\em chiral central
charge} $c$, which characterizes its gapless edge modes.  It is
directly measurable as a quantized thermal Hall conductivity,
$\kappa^{\rm q}_{xy}=\pi c T/6$ ($\hbar=k_B=1$).  A non-integer
value is an unambiguous indicator of a non-Abelian phase, and
$c=1/2$ for Ising anyons.

Stimulated by the recognition that Kitaev's anisotropic interactions
arise naturally in certain strongly spin-orbit coupled Mott
insulators \cite{jackeli2009mott,trebst2017kitaev}, mounting efforts have targeted
such systems in the laboratory.  There is now strong evidence
that Kitaev interactions are substantial in several 2d honeycomb
lattice materials \cite{winter2016challenges}:
$\alpha$-Na$_2$IrO$_3$ \cite{chun2015direct},
$\alpha$-Li$_2$IrO$_3$ \cite{williams2016incommensurate}, and
$\alpha$-RuCl$_3$ \cite{plumb2014alpha}. While it is clear that none of these materials
are exactly described by Kitaev's model, the beauty of a topological
phase is its robustness: once obtained, it is stable to an arbitrary
weak perturbation and its essential properties are completely
independent of the details of the Hamiltonian.  A very recent
experiment \cite{kasahara2018majorana} presents observations of an
apparent plateau with a quantized thermal Hall conductivity with
$c=1/2$ in $\alpha$-RuCl$_3$ in an applied field of 9-10T, at
temperatures of 3-5K.  If confirmed, it could be a revolutionary
discovery not only in the non-Abelian context, but also as the first
truly unambiguous signature of a quantum spin liquid phase in experiment. These results appear to complement recent experiments on quantum Hall systems which have observed half-integer thermal conductance, but through rather different means~\cite{Banerjee2018nonabelions}.

The $\alpha$-RuCl$_3$ experiments do, however, present at least one
major puzzle.  The thermal {\em Hall angle} $\theta_H =
\tan^{-1}(\kappa_{xy}/\kappa_{xx}) = 10^{-3}$ is small, i.e.,
$\kappa_{xx}\gg\kappa_{xy}$.  This is incompatible with conduction
solely through a Majorana edge mode.  Indeed, in two dimensional
electron gases, a quantized Hall effect is only observed when the
Hall angle is large.  This raises the fundamental question of
whether the thermal Hall effect is different: is quantization even
expected and possible at small Hall angles?  We consider here a
universal effective model for an Ising anyon phase, in which the
chiral Majorana edge mode is augmented by acoustic bulk phonons,
which can provide a diagonal bulk thermal conductivity. Remarkably,
we find that not only does the quantized thermal Hall effect persist
in the presence of the phonons, but it {\em relies} upon them.  The
ultimate view of the quantized transport is distinctly different
from  the usual isolated edge mode picture, and we predict notable
experimental consequences of the mixing of edge and bulk heat
propagation.  Our considerations are quite general and we expect
that similar physics applies to thermal transport in other systems
with edge modes, such as topological superconductors and quantum
Hall systems.

\begin{figure}[bth!]
\includegraphics[width=\columnwidth]{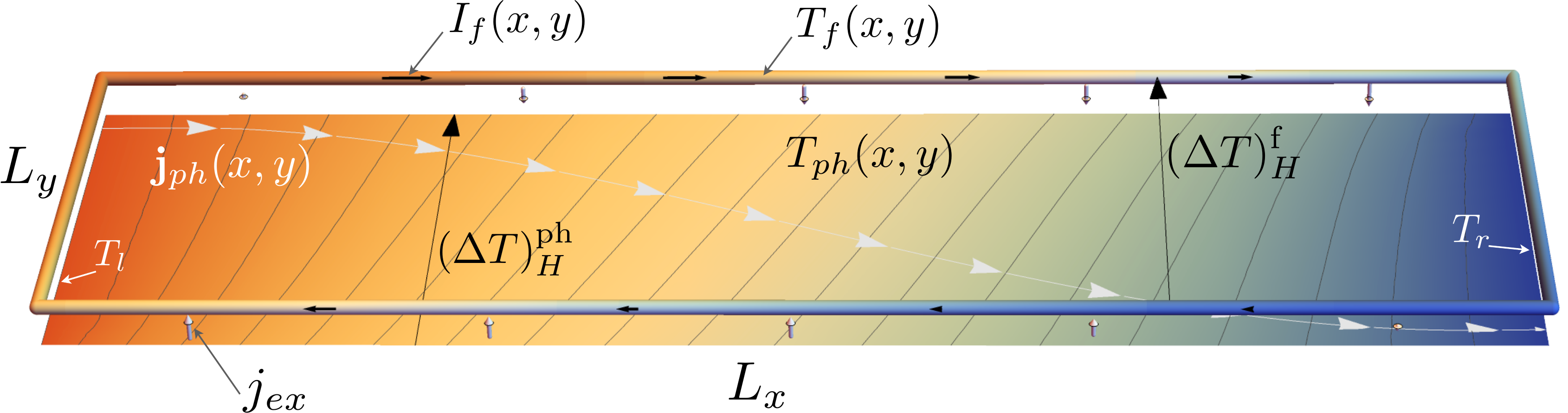}
\caption{Temperature maps of our rectangular system with dimensions $L_x$ and $L_y$ consisting
of a phonon bulk (lower box) and a Majorana fermion edge  (upper
edge). The phonon temperatures at the left and right edges are assumed
to be fixed as $T_{l,r}$, respectively, due to the coupling of the lattice with the heater and thermal bath. The black arrows for $I_f$ along the edge denote the direction and magnitude of the
``clockwise'' energy current associated with the chiral Majorana mode. The white arrows in the bulk show a stream line of $\mathbf{j}_{ph}$. The 3d white arrows for $j_{ex}$ indicate the energy current between the Majorana edge and bulk phonons. $(\Delta T)_{H}^{ph}$ and $(\Delta T)_{H}^{f}$ are the measured ``Hall" temperature differences when the contacts are coupled to the lattice or spins, respectively.
} \label{fig-1}
\end{figure}

We formulate the problem in terms of hydrodynamic equations
describing the energy transport. We consider the following two
subsystems: the phonons, or lattice, located in the bulk, and
denoted with the index ``$ph$'', and the Majorana fermions, or
spins, confined to the edge and indexed by ``$f$'', as well as a
coupling between them. For simplicity, we assume an isotropic bulk,
with the relation
\begin{equation}
  \label{eq:3}
  \mathbf{j}_{ph}=-\kappa\boldsymbol{\nabla}T_{ph},
\end{equation}
i.e., the energy current density in the bulk is parallel to the
thermal gradient, with $\kappa$ a characteristic of the lattice. The
``clockwise" edge current is that of a chiral fermion with central
charge $c=1/2$, i.e.,
\begin{equation}
  \label{eq:4}
  I_f=\frac{\pi cT_f^2}{12}.
\end{equation}
The heat exchange between the phonons and Majoranas can be modeled
phenomenologically through an energy current
$j_{ex}$ between the two subsystems (see the arrows in Fig.~\ref{fig-1}).
Microscopically, it is due to the
scattering events between edge Majorana fermions and bulk phonons,
and is the rate of energy transfer at the edge per unit length,
i.e., $j_{ex}\equiv
\frac{1}{L}\left(\frac{\partial\mathcal{E}}{\partial
t}\right)_{ph\rightarrow
f}=-\frac{1}{L}\left(\frac{\partial\mathcal{E}}{\partial
t}\right)_{f\rightarrow ph}$, where $L$ is the length of the edge in
evaluating $\left(\frac{\partial\mathcal{E}}{\partial
t}\right)_{ph\rightarrow f}$ \footnote{Note that $L=L_x$, resp.\ $L=L_y$, for $j_{ex}({\rm top/bottom})$, resp.\ $j_{ex}({\rm left/right})$}. This in turn implies that the phonons and Majoranas have not fully thermalized with one another. Assuming, however, that thermalization is almost complete, i.e., $T_f\approx T_{ph}$, and that the fermions are strictly confined to the edge, $j_{ex}$ can be linearized in the temperature difference $T_{ph}-T_f$ at the edge,
\begin{equation}
  \label{eq:def_jex1}
  j_{ex} =  \lambda(T) (T_{ph} - T_{f}),
\end{equation}
where, crucially, $\lambda>0$ is a function of the overall constant temperature
$T\approx T_{ph,f}$, and can be parametrized as $\lambda(T)\sim T^\alpha$. We will
determine $\alpha$ from a phase space analysis of the scattering events.

\emph{Hydrodynamic equations.---}We assume our (two-dimensional)
system to be a rectangular slab of width $L_y$ and length $L_x
\gtrsim L_y$ (see Fig.~\ref{fig-1}), and choose coordinates with
$|x|< x_0=L_x/2$ and $|y|<y_0=L_y/2$.

The continuity equation in the bulk in a steady state is
$\boldsymbol{\nabla}\cdot\mathbf{j}_{ph}(x,y)=0$ which implies the Laplace equation
\begin{equation}
  \label{eq:5}
\nabla^2T_{ph}(x,y)=0.
\end{equation}
Energy conservation at the edges gives rise to appropriate
boundary conditions.  At the
left and right edges, we assume that only the \textit{lattice} is
coupled to thermal leads and the phonons have fixed constant
temperatures, $T_{l,r}$, respectively. At the top and bottom edges, the current out of the phonon subsystem
must equal the exchange current, hence  $\pm j^y_{ph}(x,\pm
y_0)=j_{ex}(x,\pm y_0)$.  Moreover, the continuity equations for the
edges imply $\pm\partial_x I_f(x,\pm y_0)=j_{ex}(x,\pm y_0)$.
Together these yield, given Eqs.~\eqref{eq:3} and \eqref{eq:4},
\begin{equation}
  \label{eq:boundary_jex1}
\kappa\partial_yT_{ph}(x,\pm
    y_0)=-\kappa_{xy}^{\rm q}\partial_xT_f (x,\pm
    y_0).
\end{equation}
Note the appearance of the ideal quantized Hall conductivity
$\kappa_{xy}^q= \pi c T/6 = \pi T/12$ here, using $T_f\approx T$ (valid within our linearized treatment).

\emph{Quantization in the infinitely long limit.---}For simplicity,
we first solve our hydrodynamic equations in the limit of an
infinitely long system ($L_x \rightarrow \infty$). Note that, even
for finite systems with $L_x \gg L_y$, this infinitely long limit is
expected to be relevant far away from the left and right edges.

Since there is translation symmetry in the $x$ direction, the boundary
conditions $T_{ph}(\pm x_0,y)=T_{r,l}$ lead to a
uniform temperature gradient
$\overline{\frac{dT}{dx}}=\lim_{L_x\rightarrow\infty}\frac{T_r-T_l}{L_x}$,
and the phonon and Majorana temperatures must take the forms
\begin{equation}
  \label{eq:infty-T}
  \begin{cases}
    T_{ph}(x,y) = \overline{\frac{dT}{dx}} x + \hat{T}
(y)+{\rm const.}\\
T_f (x, \pm y_0) = \overline{\frac{dT}{dx}} x +{\rm const.}.
  \end{cases}
\end{equation}
Laplace's equation, Eq.~\eqref{eq:5}, immediately implies that
$\hat{T} (y)$ must be a linear function of $y$ which we write
$\hat{T} (y) = \frac{(\Delta T)_H^{\rm ph}}{L_y}y$. Therefore, from
Eq.~\eqref{eq:boundary_jex1}, we get
\begin{equation}
  \label{eq:10}
  \partial_yT_{ph}(x,y)=-\frac{\kappa_{xy}^{\rm q}}{\kappa}\overline{\frac{\diff T}{\diff x}},
\end{equation}
since $\partial_yT_{ph}(x,y)=\partial_yT_{ph}(x,\pm y_0)={\rm const.}$.
From a phenomenological perspective, the {\em total} current in the
Hall bar geometry must flow only along $x$, but Eq.~\eqref{eq:10}
implies that the phonon thermal gradient is tilted from the current
axis by a small Hall angle of $|\tan\theta_H |= \kappa_{xy}^{\rm q}/\kappa
\ll 1$.

Next consider the  view of Alice the experimentalist.  She measures
the temperature gradients via three contacts, and assumes for the
moment that these measurements give the phonon temperature (the most
reasonable assumption).   To deduce the Hall conductivity, she
posits a bulk heat current satisfying $\bm{j} = - \bm{\kappa}^{\rm
ph, {\rm expt}}\cdot \bm{\nabla} T$, and tries to deduce the tensor
$\bm{\kappa}^{\rm ph, {\rm expt}}$ [the $ph$ ($f$) superscript means this quantity is obtained from a measurement of the phonon (Majorana fermion) temperature]. By measuring the longitudinal temperature gradient, she obtains $\kappa^{\rm ph,{\rm expt}}_{xx} = \kappa$ as expected, and then, imposing $j^y=0$, she equates the experimental Hall angle $\tan\theta_H = \frac{(\Delta T)^{\rm ph}_H}{L_y} /
\overline{\frac{dT}{dx}}$ to $\kappa_{xy}^{\rm ph,{\rm expt}} /
\kappa^{\rm ph,{\rm expt}}_{xx}$. By comparing this equation to
the theoretical result in Eq.~\eqref{eq:10}, we immediately
recognize that the magnitude of the effective Hall conductivity (denoted simply as $\kappa_{xy}^{\rm expt}$ in the rest of the text) is $|\kappa_{xy}^{\rm ph, expt}|=\kappa_{xy}^{\rm q}$, i.e., the experimentally measured thermal
Hall conductivity takes the quantized value!

\begin{figure}
\includegraphics[width=1.0\columnwidth]{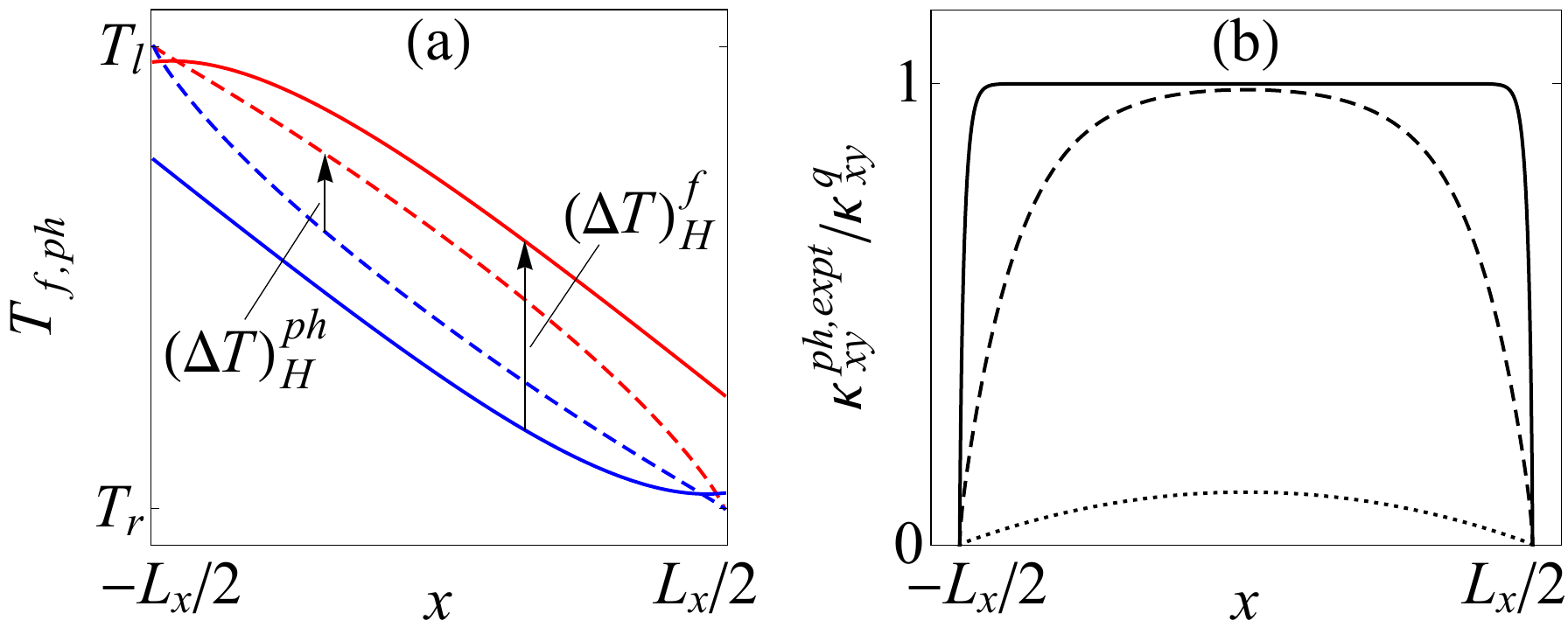}
\caption{(a) Temperature profiles of the
    Majorana fermions (solid lines) and phonons (dashed lines) at the
    top (red lines) and bottom (blue lines) edges, $T_{f,ph}(x,\pm
      y_0)$. The measured ``Hall" temperature differences $(\Delta
      T)_{H}^{ph,f}(x)\equiv T_{ph,f}(x,y_0)-T_{ph,f}(x,-y_0)$ are shown
      with the black arrows. 
(b) Measured thermal Hall conductivity $\kappa_{xy}^{\rm ph,expt}$
[Eq.~\eqref{eq-kappa-3}] as a function of the longitudinal position
$x$ at which $(\Delta T)_{H}^{ph}$ is measured for dimensionless
thermal couplings $\lambda L_x / \kappa_{xy}^{\rm q} = 100$ (solid
line), $10$ (dashed line), and $1$
(dotted line) at fixed $L_x / L_y = 100$.} 
\label{fig-2}
\end{figure}

A few remarks are in order. 
First, a transverse temperature difference, $(\Delta T)^{\rm ph}_H$,
leading to a ``Hall thermal gradient'' $(\Delta T)^{\rm
  ph}_H/L_y=-\frac{\kappa_{xy}^{\rm q}}{\kappa}\overline{\frac{dT}{dx}}$
develops which allows to compensate the transverse energy current
$j_{ex}$ at the edges and leads to a zero net transverse current.
Second, the effective thermal Hall conductivity
is only found to be quantized if the transverse temperature gradient
is obtained from the \emph{phonon} temperatures at the top and
bottom edges. In contrast, if Bob somehow measures the Majorana
temperatures, the transverse temperature gradient is identified as
$(\Delta T)^{\rm f}_H/L_y$ and thus, from Eqs.~\eqref{eq:def_jex1}
and~\eqref{eq:boundary_jex1}, he finds a different effective thermal
Hall conductivity [see also Fig.~\ref{fig-2}(a)]:
\begin{equation}
\kappa_{xy}^{\rm f, expt} = -\frac{ \kappa (\Delta T)^{\rm f}_H} {L_y \overline{\frac{dT}{dx}}} =
\kappa_{xy}^q \left( 1 + \frac{2 \kappa} {\lambda(T) L_y}
\right). \label{eq-kappa-2}
\end{equation}
Note that $\kappa_{xy}^{\rm f,{\rm expt}} \approx \kappa^{\rm ph,{\rm expt}}_{xy}$ only for a large enough phonon-Majorana coupling
$\lambda(T) \gg \kappa / L_y$.

\emph{General conditions for quantization.---}To understand how the
quantization of the effective thermal Hall conductivity can break
down and determine the range of its applicability, we now extend the
solution of our hydrodynamic equations to a finite system with $L_x
\gtrsim L_y$, where we must take into account all boundary
conditions, i.e., include the right and left boundary conditions on
top of those in Eq.~\eqref{eq:boundary_jex1}. Again assuming that
the leads are coupled to the phonons only, those are:
\begin{equation}
  \label{eq:boundary_jex2}
  \begin{cases}
    T_{ph}(\pm x_0,y)=T_{r,l},\\
j_{ex}(\pm x_0,y)= \lambda(T) (T_{ph} - T_{f})=\mp \kappa_{xy}^{\rm
q}\partial_yT_f.
  \end{cases}
\end{equation}
Considering a small enough phonon-Majorana coupling $\lambda$, we
aim to obtain a perturbative solution of the hydrodynamic equations.
To this end, we write
\begin{equation}
  \label{eq:12}
  T_{ph,f}(x,y)={T} + \tilde T_{ph,f}(x,y),
\end{equation}
with $\tilde T_{ph,f}(x,y) \ll {T}$. We express the temperature variations in series
expansions as $\tilde T_{ph,f} = \sum_{n=0}^{\infty} \tilde
T_{ph,f}^{(n)}$ and assume that terms of
increasing order $n$ are progressively less important. Note also that $\tilde T_{ph,f}(x,y) = -\tilde T_{ph,f}(-x,-y)$ generally follows from the symmetries of the
hydrodynamic equations. Starting from
the $\lambda = 0$ solution, $\tilde T_{ph}^{(0)}(x,y) =\overline{\frac{dT}{dx}} x$ and
$\tilde T_f^{(0)}(x,y) = 0$, the temperature variations can then be found
by an iterative procedure. At each iteration step $n > 0$, we first
solve the ordinary differential equations [see
Eqs.~\eqref{eq:boundary_jex1} and \eqref{eq:boundary_jex2}]
\begin{eqnarray}
& \kappa_{xy}^{\rm q} \partial_x \tilde T_f^{(n)} =
\pm \lambda \left[ \tilde T_{ph}^{(n-1)} - \tilde T_f^{(n)} \right]\quad &\textrm{for}\; y = \pm y_0,
\nonumber \\
& \kappa_{xy}^{\rm q} \partial_y \tilde T_f^{(n)}
= \mp \lambda \left[ \tilde T_{ph}^{(n-1)} - \tilde T_f^{(n)} \right] \quad &  \textrm{for}\; x = \pm x_0,
\label{eq-it-1}
\end{eqnarray}
for the Majorana temperature $\tilde T_f^{(n)}$ along the edge. Then,
using this solution, we obtain an appropriate Laplace equation
$\nabla^2 \tilde T_{ph}^{(n)} = 0$ for the phonon temperature $\tilde T_{ph}^{(n)}$ in the
bulk, along with Dirichlet boundary conditions $\tilde T_{ph}^{(n)} (\pm x_0, y)
= 0$ at the left and right edges, and Neumann boundary conditions
\begin{equation}
\partial_y \tilde T_{ph}^{(n)} = \pm \frac{\lambda}
{\kappa} \left[ \tilde T_f^{(n)} - \tilde T_{ph}^{(n-1)} \right] \qquad  \textrm{for}\; y = \pm y_0,
\label{eq-it-2}
\end{equation}
at the top and bottom edges. It is well known that such a Laplace
equation with mixed Dirichlet and Neumann boundary conditions has a
unique solution that can be obtained by standard methods. Our perturbative solution is convergent whenever $\lambda\ll \kappa / L_y$ (see~\cite{suppmat} for the error analysis).

Assuming this condition, we perform the first iteration step (see
\cite{suppmat}) to calculate the phonon temperature $\tilde
T_{ph}^{(1)}$ and obtain the effective thermal Hall conductivity in
terms of the transverse temperature difference $(\Delta T)_{H}^{\rm
ph}(x)$ [see Fig.~\ref{fig-2}(a)]:
\begin{equation}
\kappa_{xy}^{\rm ph,expt} (x) =- \frac{\kappa} {\overline{\frac{dT}{dx}} L_y} \left[
  \tilde T_{ph}^{(1)} (x,
y_0) - \tilde T_{ph}^{(1)} (x, -y_0) \right]. \label{eq-kappa-3}
\end{equation}
Note that $\kappa_{xy}^{\rm ph,expt} (x)$ generally depends on
the position $x$ at which the temperatures are measured [see
Fig.~\ref{fig-2}(b)]. Indeed, we find that $\kappa_{xy}^{\rm
ph,expt} (x)$ only takes a quantized (or even constant) value if
$L_x\gg L_y$ and $L_x\gg \ell \equiv \kappa_{xy}^{\rm q} / \lambda$.
First, an accurate measurement of the thermal Hall conductivity
generally requires an elongated system with $L_x \gg L_y$. Second,
the system size $L_x$ must be larger than the characteristic length
$\ell$ associated with the thermalization of the Majorana edge mode
(see Table~\ref{tab1} for a summary). Indeed, even for $L_x \gg
L_y$, there are two regimes for the effective thermal Hall
conductivity (see \cite{suppmat}):
\begin{eqnarray}
\kappa_{xy}^{\rm ph,expt} (x) \approx   \left\{\begin{array}{cc}
\frac{\pi {T}} {12} & (L_x \gg \ell), \\
\frac{\pi {T} (L_x^2 - 4x^2)}{96\ell^2} & (L_x \ll \ell).
\end{array}\right.
\label{eq-kappa-4}
\end{eqnarray}
In the second regime we find that $\kappa_{xy}^{\rm ph,expt} (x)$ has a strong dependence on $x$ and is smaller than $\kappa_{xy}^{\rm q} = (\pi /12) {T}$ by a factor $\sim (L_x / \ell)^2 \ll 1$.

\emph{Estimation of the spin-lattice thermal coupling $\lambda$.---}
The phenomenological spin-lattice coupling $\lambda(T)$ defined in
Eq.~\eqref{eq:def_jex1} can be obtained microscopically from, 
e.g., the Boltzmann equation. We calculate the rate
of energy exchange per unit length
$j_{ex}=\frac{1}{L}\Big(\frac{\partial \mathcal{E}}{\partial
t}\Big)_{ph\rightarrow f}$ due to the 
scattering at
the edge. Comparing to the form in Eq.~\eqref{eq:def_jex1}, we
extract $\lambda(T) = \lambda_0 T^\alpha$, i.e., the exponent
$\alpha$ and the coefficient $\lambda_0$.

We consider a coupling at the top edge
$y=y_0=L_y/2$ of the form
\begin{align}\label{eq:coupling}
H_{int}=\frac{- i g v_f}{4}\int \diff  x \, \zeta(
  x)K_{ij}\partial_{i} u_{j}(x,y_0) \eta( x)\partial_{ x}\eta( x),
\end{align}
where $\eta(x),\,\vec{u}(x,y), \zeta(x)$ are the Majorana edge mode,
the lattice displacement field, and disorder potential,
respectively, $g$ parametrizes the spin-lattice coupling, and $v_f$
is the fermion velocity. $K_{ij}\partial_{i}u_{j}$ with $i,j=x,y$ is some linear
combination of the elastic tensor for $u$. Physically, Eq.~\eqref{eq:coupling} may be understood from the
observation that the lattice displacement modifies the velocity of the
Majorana edge mode by affecting the strength of the Kitaev coupling.

Using Eq.~\eqref{eq:coupling} and calculating the energy transfer rate
using a Boltzmann equation, we obtain a large power $\alpha=6$.
The reason for the large exponent is twofold.  First, the dispersions
of both bulk phonons and edge Majoranas are linear which reduces the
low energy phase space. Second, the vertex necessarily involves two 
gradients: one because $\eta(x)\eta(x)=\delta(0)$ is a c-number
for Majorana fermions, and another because the strain tensor includes 
a gradient. We note that, without disorder, two-phonon processes are necessary to satisfy kinematic constraints in the physical regime, where the velocity of the acoustic phonon $v_{ph}$ is larger than $v_f$.
In that case one obtains an even larger $\alpha=8$.

To estimate the coefficient $\lambda_0$, we further assume that the
averaged disorder potential satisfies $\la \zeta(x)\zeta(x')
\ra_{dis}=\zeta^2\,\delta(x-x')$, and consider an isotropic acoustic
phonon mode only. From the Boltzmann equation solution
(see \cite{suppmat}), we obtain
\begin{align}\label{eq:current0}
\lambda=\frac{g^2\zeta^2}{32(2\pi)^3v_{ph}^4 v_f^2\rho_0}f \, {T}^6,
\end{align}
where $\rho_0$ is the mass density of the lattice. In the model we
consider, $f=4.2\times10^4$.
Unfortunately, at this time an accurate quantitative estimate of $\lambda$ for $\alpha$-RuCl$_3$
is not possible due to the lack of knowledge of microscopic details of
$g$, $v_f$ and $\zeta$. 
However, crudely applying Eq.~\eqref{eq:current0}, we
estimate the characteristic length $\ell = \kappa_{xy}^{\rm q}/\lambda$ to be
several orders of magnitude larger than the lattice spacing at
temperatures of a few Kelvins.   Importantly, due to the large
 exponent $\alpha$, we expect that upon lowering the temperature of the
sample, $\ell$ grows rapidly and that the system 
enters 
the regime where 
$L_x\ll \ell$ in Eq.~\eqref{eq-kappa-4} and thus the quantization of
the thermal Hall conductivity breaks down.

\emph{Summary and discussion.---}By carefully analyzing the
interplay between the chiral Majorana edge mode of an Ising anyon
phase and the energy currents carried by bulk phonons, we have
demonstrated that the thermal Hall conductivity of such a
non-Abelian topological phase can be effectively quantized in the
presence of a much larger longitudinal thermal conductivity. This 
is in accordance with recent experiments on $\alpha$-RuCl$_3$
\cite{kasahara2018majorana}. However, this quantization only
survives under certain conditions. The main results are summarized in Table~\ref{tab1}.

\begin{table}[h]
\begin{tabular}{>{\centering}p{1.2cm}>{\centering}p{2.1cm}>{\centering}p{2.1cm}>{\centering}p{2.1cm} }
\hline\hline
\multicolumn{1}{c|}{Coupling regime}          &  Weak  &
Intermediate & Strong  \tabularnewline
\multicolumn{1}{c|}{$\lambda\sim T^{\alpha}$}  & $\lambda\lesssim
\lambda_f$  &
                                                           $\lambda_f\ll\lambda\ll \lambda_{ph}$ & $\lambda_{ph}\ll\lambda$  \tabularnewline
\multicolumn{1}{c|}{$L_x$}         & {$L_x\lesssim \ell$} & \multicolumn{2}{c}{$L_x\gg \ell$}  \tabularnewline
\multicolumn{1}{c|}{$L_y$}          & \multicolumn{2}{c}{$L_y\ll
                                       \kappa/\lambda$} & {$L_y\gg\kappa/\lambda$}
  \tabularnewline
\hline
\multicolumn{1}{c|}{$\kappa_{xy}^{\rm ph,expt}$} & {$\kappa_{xy}^{\rm ph,expt}\ll\kappa_{xy}^{\rm q}$} &
{$\kappa_{xy}^{\rm q}$} &
{$\kappa_{xy}^{\rm q}$}  \tabularnewline
\multicolumn{1}{c|}{$\kappa_{xy}^{\rm f,expt}$}& --$^\text{\cite{tablenote}}$ & $\kappa_{xy}^{\rm f,expt}\gg \kappa_{xy}^{\rm q}$ & {$\kappa_{xy}^{\rm q}$} \tabularnewline
\hline\hline
\end{tabular}
\caption{Values of the effective thermal Hall conductivities extracted by measuring the temperatures of the phonon ($\kappa_{xy}^{\rm ph,expt}$) or Majorana ($\kappa_{xy}^{\rm f,expt}$) subsystems in three coupling regimes, defined by the value of $\lambda$ relative to $\lambda_f=\kappa_{xy}^{\rm q} / L_x$ and $\lambda_{ph}=\kappa / L_y$. The three coupling regimes can also be identified by comparing the system dimensions $L_x, \, L_y$ to the characteristic lengths $\ell =\kappa_{xy}^{\rm q} / \lambda$ and $\kappa/\lambda$.}
\label{tab1}
\end{table}

In words, those results are as follows. The quantization survives for a sufficiently strong spin-lattice coupling $\lambda
\gg \lambda_f \equiv \kappa_{xy}^{\rm q} / L_x$, while it immediately
disappears in the weak-coupling regime defined by $\lambda \lesssim \lambda_f$ [see Fig.~\ref{fig-2}(b)]. Importantly, since $\lambda
\propto T^{\alpha}$ is strongly dependent on the temperature, with
$\alpha \geq 6$ for the mechanisms considered in this work, we
predict that the observed quantization of the thermal Hall
conductivity should eventually break down as the temperature is
lowered.

Even within the range of quantization ($\lambda \gg \lambda_f$), we
can identify two separate regimes, depending on how $\lambda$
compares to $\lambda_{ph} \equiv \kappa / L_y \gg \lambda_f$. In
the strong-coupling regime, defined by $\lambda \gg \lambda_{ph}$, the spins and the lattice share the same temperature, and the quantization of the thermal Hall conductivity follows from effectively having a
  system with a diagonal conductivity $\kappa_{xx}^{\rm expt}=\kappa_{yy}^{\rm expt}=\kappa$
of the phonons and an off-diagonal $\kappa_{xy}^{\rm
  expt}=\kappa_{xy}^q$ of the Majoranas. Surprisingly, however, in the intermediate
regime defined by 
$\lambda_f \ll \lambda \ll \lambda_{ph}$, the thermal Hall
conductivity appears to be quantized \emph{despite} a large
temperature mismatch between the spins and the lattice. This is only
true, however, if it is obtained by measuring the \emph{lattice}
temperatures along the edge. If one could directly measure the local
temperature of the Majorana edge mode, it would appear to give a
much larger thermal Hall conductivity.

Finally, we emphasize that our hydrodynamic equations are applicable
far beyond the scope of the present work. Here, by solving them, we
obtained a wide range of experimentally measurable quantities, such
as detailed temperature profiles of various degrees of freedom
(e.g., spins and lattice) across the system. However, due to their
phenomenological nature, the hydrodynamic equations we derived
should readily extend to a rich variety of chiral topological phases
and thus may find applications far away from the field of quantum
spin liquids.


\begin{acknowledgments}
L. B. was supported by the DOE Office of Science's Basic Energy
  Sciences program under Award No.\ DE-FG02- 08ER46524.  M. Y.
  acknowledges support from the KITP graduate fellowship program under Grant No. NSF PHY-1748958 and the NSF DMR-1523036 from the University of Minnesota. G. B. H. is supported by the Gordon and Betty Moore Foundation's EPiQS Initiative through
  Grant No.~GBMF4304.
\end{acknowledgments}

\bibliography{majorana-thermal-hall.bib}

\begin{thebibliography}{22}
\bibitem [{tab()}]{tablenote}%
  \BibitemOpen
\bibfield  {journal} {  }\href@noop {} {\bibinfo  {journal} {{``--" in the last
  line means that the quantization $\kappa_{xy}^{\rm f,expt}$ relative to
  $\kappa_{xy}^{\rm q}$ is not generic ($\kappa_{xy}^{\rm f,expt}\gtrless
  \kappa_{xy}^{\rm q}$) in the weak coupling regime $\lambda\lesssim\lambda_f$,
  i.e., it depends on the strength of $\lambda$ and the position $x$ where the
  temperature is measured}}\ }\BibitemShut {NoStop}%
\end{thebibliography}
\onecolumngrid
\clearpage
\appendix
\section*{Supplementary Material}

\section{First-order solution of the hydrodynamic equations}
\label{sec:first-order-solution}

Here we work out the first iteration step of the perturbative
solution described in the main text and use it to determine the
effective thermal Hall conductivity $\kappa_{xy}^{\rm ph,expt} (x)$ as a function
of the longitudinal position $x$. At the first iteration step ($n =
1$), the ordinary differential equations for the Majorana
temperature $\tilde T_f^{(1)} (x,y)$ in Eq.~(11) of the main text are
\begin{eqnarray}
\partial_x \tilde T_f^{(1)} \left( x, \pm \frac{L_y}{2} \right) &=& \pm
\frac{1} {\ell} \left[ \tilde T_{ph}^{(0)} \left( x, \pm \frac{L_y}{2} \right) -
\tilde T_f^{(1)} \left( x, \pm \frac{L_y}{2} \right) \right],
\nonumber \\
\nonumber \\
\partial_y \tilde T_f^{(1)} \left( \pm \frac{L_x}{2}, y \right) &=& \mp
\frac{1} {\ell} \left[ \tilde T_{ph}^{(0)} \left( \pm \frac{L_x}{2}, y \right) -
\tilde T_f^{(1)} \left( \pm \frac{L_x}{2}, y \right) \right],
\label{eq-hydro-ord}
\end{eqnarray}
where $\ell = \kappa_{xy}^{\rm q} / \lambda$ is the characteristic
thermalization length of the Majorana edge mode. Remembering that
$\tilde T_{ph}^{(0)} (x,y) = \overline{\frac{dT}{dx}} x$, 
the solutions of
Eq.~\eqref{eq-hydro-ord} at the left and top edges become
\begin{eqnarray}
\tilde T_f^{(1)} \left( -\frac{L_x}{2}, y \right) &=&
-\overline{\frac{dT}{dx}} \left\{ \frac{L_x}{2} - \frac{\ell [1 -
e^{-L_x / \ell}]} {1 + e^{-(L_x + L_y) / \ell}} \, \exp \left[
-\frac{y + L_y/2}{\ell} \right] \right\},
\label{eq-hydro-ord-sol} \\
\nonumber \\
\tilde T_f^{(1)} \left( x, \frac{L_y}{2} \right) &=&
-\overline{\frac{dT}{dx}} \left\{ (\ell - x) - \frac{\ell [1 +
e^{-L_y / \ell}]} {1 + e^{-(L_x + L_y) / \ell}} \, \exp \left[
-\frac{x + L_x/2}{\ell} \right] \right\}, \nonumber
\end{eqnarray}
while those at the right and bottom edges are related by the
symmetry constraint $\tilde T_f^{(1)} (x,y) = -\tilde T_f^{(1)}
(-x,-y)$. Laplace's equation $\nabla^2 \tilde T_{ph}^{(1)} (x,y) =
0$ for the phonon temperature $\tilde T_{ph}^{(1)} (x,y)$ is then
supplemented with the Dirichlet boundary conditions $\tilde
T_{ph}^{(1)} (\pm L_x / 2, y) = 0$ at the left and right edges and
the Neumann boundary conditions [see Eq.~(12) in the main
text]
\begin{eqnarray}
\partial_y \tilde T_{ph}^{(1)} \left( x, \pm \frac{L_y}{2} \right) &=& \pm
\frac{\lambda} {\kappa} \left[ \tilde T_f^{(1)} \left( x, \pm
\frac{L_y}{2} \right) - \tilde T_{ph}^{(0)} \left( x, \pm
\frac{L_y}{2} \right)
\right] \nonumber \\
\nonumber \\
&=& \frac{-\overline{\frac{dT}{dx}} \kappa_{xy}^{\rm q}} {\kappa} \left[ 1 - \frac{1 +
e^{-L_y / \ell}} {1 + e^{-(L_x + L_y) / \ell}} \, \exp \left( -\frac{L_x/2
\pm x}{\ell} \right) \right] \label{eq-hydro-bound-1}
\end{eqnarray}
at the top and bottom edges. It is well known that Laplace's
equation with such boundary conditions has a unique solution. After the iteration step $n=1$, the only error in the temperature corrections
$\tilde T_{ph}^{(1)}$ and $\tilde T_f^{(1)}$ is due to the absence of
$\tilde T_{ph}^{(1)}$ on the right-hand sides of Eqs.~\eqref{eq-hydro-ord} and \eqref{eq-hydro-bound-1}.
Indeed, including this term would precisely give rise to the next
temperature corrections $\tilde T_{ph}^{(2)}$ and $\tilde T_f^{(2)}$. The same type of error from $\tilde T_{ph}^{(n)}$ persists in the same way at a given iterative step $n$. However, it follows from Eq.~\eqref{eq-hydro-bound-1} [see also Eq.~(12)] that successive temperature corrections $\tilde T_{ph}^{(n)}$ are progressively less important
and hence our perturbative solution is convergent whenever $\lambda
\ll \kappa / L_y$. Surprisingly, as long as $\kappa_{xy}^{\rm q}
\ll \kappa$, the perturbative solution is actually valid for an
\emph{arbitrary} value of $\lambda$. Indeed, at $\lambda \rightarrow
\infty$, one can formulate an equivalent perturbative solution for
$\tilde T_{ph} = \tilde T_f$ at the edges in which the small parameter is manifestly $\kappa_{xy}^{\rm q} / \kappa$.
 
Recalling the symmetry constraint $\tilde T_{ph}^{(1)} (x,y) = -\tilde
T_{ph}^{(1)}
(-x,-y)$, we search for this solution in the general form
\begin{equation}
\tilde T_{ph}^{(1)} (x,y) = \sum_{m = 1}^{\infty} \bigg\{ A_m \cos \left[
\frac{(2m-1) \pi x} {L_x} \right] \sinh \left[ \frac{(2m-1) \pi y} {L_x}
\right] + B_m \, \sin \left[ \frac{2m \pi x} {L_x} \right] \cosh
\left[ \frac{2m \pi y} {L_x} \right] \bigg\}, \label{eq-hydro-T}
\end{equation}
which automatically satisfies the Dirichlet boundary conditions. To
find independent equations for the coefficients $A_m$ and $B_m$, we
take symmetric and antisymmetric combinations of the Neumann
boundary conditions in Eq.~\eqref{eq-hydro-bound-1}:
\begin{eqnarray}
\partial_y \tilde T_{ph}^{(1)} \left( x, \frac{L_y}{2} \right) + \partial_y
  \tilde T_{ph}^{(1)}
\left( x, -\frac{L_y}{2} \right) &=& \frac{-2
\overline{\frac{dT}{dx}} \kappa_{xy}^{\rm q}} {\kappa} \left\{ 1 -
\frac{1 + e^{-L_y / \ell}} {1 + e^{-(L_x + L_y) / \ell}} \, \exp
\left( -\frac{L_x} {2 \ell} \right) \cosh \frac{x}
{\ell} \right\} \nonumber \\
\nonumber \\
&=& \sum_{m = 1}^{\infty} \frac{2 (2m-1) \pi A_m} {L_x} \, \cos \left[
\frac{(2m-1) \pi x} {L_x} \right] \cosh \left[
\frac{(2m-1) \pi L_y} {2L_x} \right], \nonumber \\
\nonumber \\
\partial_y \tilde T_{ph}^{(1)} \left( x, \frac{L_y}{2} \right) - \partial_y
  \tilde T_{ph}^{(1)}
\left( x, -\frac{L_y}{2} \right) &=& \frac{-2
\overline{\frac{dT}{dx}} \kappa_{xy}^{\rm q} [1 + e^{-L_y / \ell}]}
{\kappa [1 + e^{-(L_x + L_y) / \ell}]} \, \exp \left( -\frac{L_x} {2
\ell} \right) \sinh \frac{x} {\ell}
\label{eq-hydro-bound-2} \\
\nonumber \\
&=& \sum_{m = 1}^{\infty} \frac{4m \pi B_m} {L_x} \, \sin \left[
\frac{2m \pi x} {L_x} \right] \sinh \left[ \frac{m \pi L_y} {L_x} \right].
\nonumber
\end{eqnarray}
From Eq.~\eqref{eq-hydro-bound-2}, the individual coefficients $A_m$
and $B_m$ in Eq.~\eqref{eq-hydro-T} are
\begin{eqnarray}
&& A_m = \frac{-\overline{\frac{dT}{dx}} L_x \kappa_{xy}^{\rm q}} {(2m-1) \pi \kappa} \,
\mathrm{sech} \left[ \frac{(2m-1) \pi
L_y} {2L_x} \right] \mathcal{I}_m, \label{eq-hydro-A} \\
\nonumber \\
&& B_m = \frac{-\overline{\frac{dT}{dx}} L_x \kappa_{xy}^{\rm q}} {2m \pi \kappa} \,
\mathrm{cosech} \left[ \frac{m \pi L_y} {L_x} \right] \mathcal{J}_m, \nonumber
\end{eqnarray}
where the Fourier integrals $\mathcal{I}_m$ and $\mathcal{J}_m$ take the forms
\begin{eqnarray}
&& \mathcal{I}_m = \frac{2}{L_x} \int_{-L_x/2}^{L_x/2} d\tilde{x} \cos \left[
\frac{(2m-1) \pi \tilde{x}} {L_x} \right] \left\{ 1 - \frac{1 + e^{-L_y
/ \ell}} {1 + e^{-(L_x + L_y) / \ell}} \, \exp \left( -\frac{L_x} {2 \ell}
\right) \cosh \frac{\tilde{x}} {\ell} \right\}, \nonumber \\
\nonumber \\
&& \mathcal{J}_m = \frac{2}{L_x} \int_{-L_x/2}^{L_x/2} d\tilde{x} \sin \left[
\frac{2m \pi \tilde{x}} {L_x} \right] \frac{1 + e^{-L_y / \ell}} {1 +
e^{-(L_x + L_y) / \ell}} \, \exp \left( -\frac{L_x} {2 \ell} \right) \sinh
\frac{\tilde{x}} {\ell}. \label{eq-hydro-I}
\end{eqnarray}
From Eqs.~\eqref{eq-hydro-T} and \eqref{eq-hydro-A}, the effective
thermal Hall conductivity in Eq.~(13) of the main text is then
\begin{eqnarray}
\kappa_{xy}^{\rm ph,expt} (x) &=& \frac{\kappa} {-\overline{\frac{dT}{dx}} L_y} \left[
                             \tilde T_{ph}^{(1)}
\left( x, \frac{L_y}{2} \right) - \tilde T_{ph}^{(1)} \left( x, -\frac{L_y}{2}
\right) \right] \label{eq-hydro-kappa-1} \\
\nonumber \\
&=& \sum_{m = 1}^{\infty} \frac{2L_x \kappa_{xy}^{\rm q}} {(2m-1) \pi L_y} \,
\mathrm{tanh} \left[ \frac{(2m-1) \pi L_y} {2L} \right] \cos \left[
\frac{(2m-1) \pi x} {L_x} \right] \mathcal{I}_m. \nonumber
\end{eqnarray}
In the limit of $L_x \gg L_y$, we can expand the function $\tanh [(2m-1)
\pi L_y / 2L_x]$ up to first order in $L_y / L_x$ and use the completeness
of the Fourier functions $\cos [(2m-1) \pi x / L_x]$ to obtain a more
tractable (but approximate) expression:
\begin{equation}
\kappa_{xy}^{\rm ph,expt} (x) = \kappa_{xy}^{\rm q} \sum_{m = 1}^{\infty} \cos \left[
\frac{(2m-1) \pi x} {L} \right] \mathcal{I}_m = \kappa_{xy}^{\rm q} \left\{ 1 -
\frac{1 + e^{-L_y / \ell}} {1 + e^{-(L_x +L_y) / \ell}} \, \exp \left(
-\frac{L_x} {2 \ell} \right) \cosh \frac{x}{\ell} \right\}.
\label{eq-hydro-kappa-2}
\end{equation}
Moreover, we can simplify $\kappa_{xy}^{\rm ph,expt} (x)$ even further by
considering the two opposite limits $L_x \gg \ell$ and $L_x \ll \ell$.
In the limit of $L_x\gg \ell$, the second term in the curly brackets
of Eq.~\eqref{eq-hydro-kappa-2} is negligible, and we find a
constant thermal Hall conductivity:
\begin{equation}
\kappa_{xy}^{\rm ph,expt} (x) = \kappa_{xy}^{\rm q}. \label{eq-hydro-kappa-3}
\end{equation}
Conversely, in the limit of $L_x \ll \ell$, we can expand the terms in
the curly brackets of Eq.~\eqref{eq-hydro-kappa-2} up to second
order in $x / \ell$ and $L_x / \ell$ (while neglecting $L_y / \ell$) to
obtain a strongly $x$-dependent thermal Hall conductivity:
\begin{equation}
\kappa_{xy}^{\rm ph,expt} (x) = \frac{\kappa_{xy}^{\rm q} (L_x^2 - 4 x^2)} {8 \ell^2}.
\label{eq-hydro-kappa-4}
\end{equation}
Finally, we recover $\kappa_{xy}^{\rm ph,expt} (x)$ in Eq.~(14) of the main text
by substituting $\kappa_{xy}^{\rm q} = (\pi / 12) {T}$ into
Eqs.~\eqref{eq-hydro-kappa-3} and \eqref{eq-hydro-kappa-4}.

\section{Microscopic calculation of the thermal coupling $\lambda$}
\label{sec:microscopic-calculation}

\begin{figure}
\includegraphics[width=0.7\columnwidth]{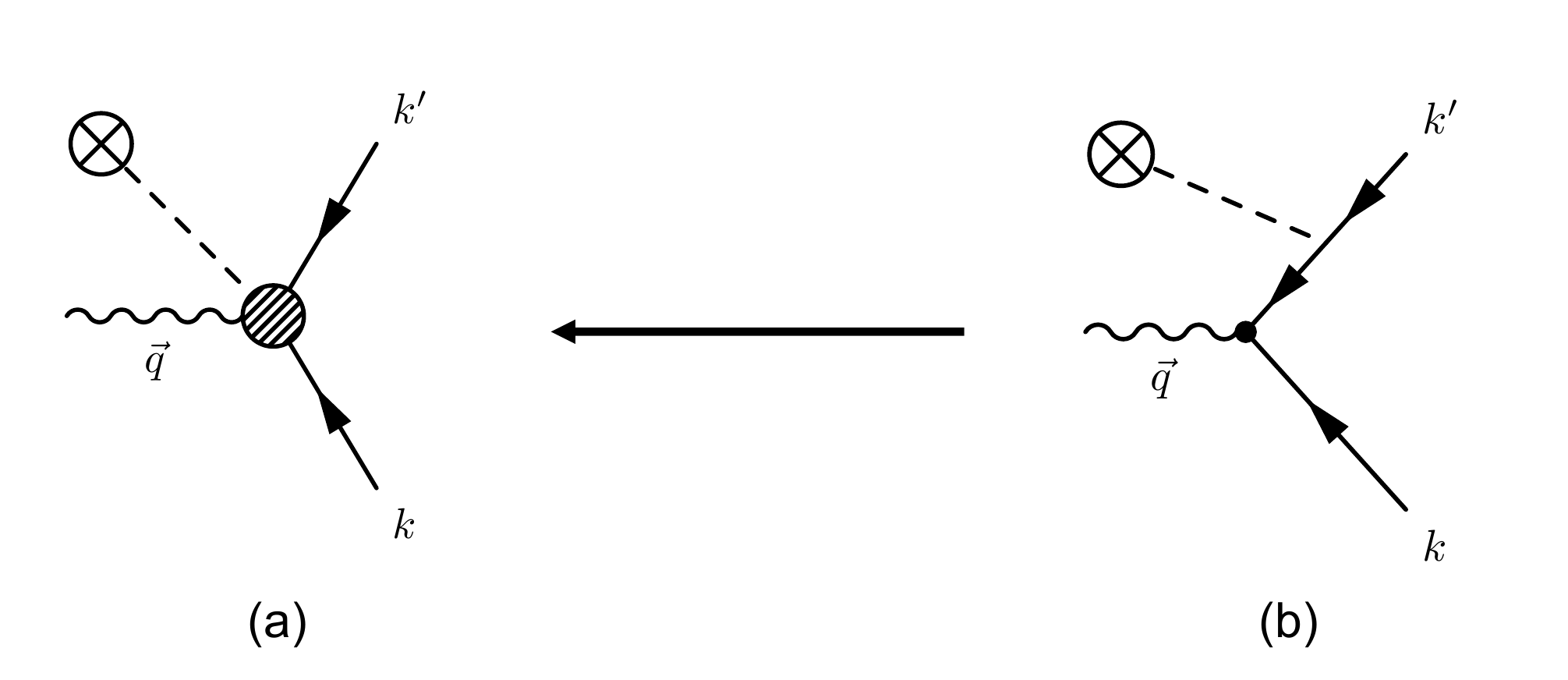}
\caption{Diagrammatic representation of the Majorana fermion (solid line) and phonon (wavy line) scattering vertex in the presence of disorder (dashed line with $\otimes$). (a) The effective Majorana-phonon-disorder vertex presented in Eq.~(15). (b) The scattering vertex constructed from the microscopic Majorana-phonon and Majorana-disorder couplings. The connection between the two is shown in Sec.~\ref{sec:EffectiveVertex}. } \label{fig-3}
\end{figure}

In this section, we start with deriving the linearized energy current between phonons and Majoranas using the Boltzmann equation formalism in Sec.~\ref{sec:collision-integral}.
In Sec.~\ref{sec:EffectiveVertex}, we justify that the Majorana-phonon-disorder vertex presented in Eq.~(15) (see Fig.~\ref{fig-3} (a)) of the manuscript can be considered as an effective vertex from the microscopic Majorana-phonon and Majorana-disorder couplings (see Fig.~\ref{fig-3} (b)).

\subsection{Collision integral}
\label{sec:collision-integral}

We consider the non-interacting Hamiltonians of the Majorana fermion and phonon fields:
\begin{align}
\mc{H}_0=\frac{-i v_f}{4}\int \diff x \,\eta(x)\partial_x\eta(x)+\frac{1}{2}\int \diff x\diff y [\rho_0^{-1}\vec{\pi}^2+B (\vec\nabla\cdot \vec{u})^2],
\end{align}
where $\vec u$ is the lattice displacement field in the continuous
limit, $\vec{\pi}$ is the conjugate lattice momentum, $\rho_0$ is the lattice mass density. The longitudinal phonon field in second quantized form is~\cite{ashcroft1976solid}:
\begin{align}
\vec{u}(\vec x, t)&=\sum_{\vec q}\left(\frac{\hbar}{2\rho_0 V
                    \omega_{\vec q}}\right)^{1/2}\hat{q}\left(c_{\vec
                    q}\,e^{i \vec q\cdot \vec x- i \omega_q
                    t}+\cd_{\vec q}\,e^{-i \vec q \cdot \vec x+i
                    \omega_{\vec q} t}\right)\non\\
\vec{\pi}(\vec x, t)&=\rho_0 \frac{\partial \vec u}{\partial
                      t}=-i\sqrt{\rho_0}\sum_{\vec q}\left(\frac{\hbar
                      \omega_{\vec q}}{2 V
                      }\right)^{1/2}\hat{q}\left(c_{\vec q}\,e^{i \vec
                      q\cdot \vec x -i \omega_{\vec q} t}-\cd_{\vec q}\,e^{-i
                      \vec q \cdot \vec x+i \omega_{\vec q} t}\right),
\end{align}
where $V=L_xL_y$ is the volume of the system, $\hbar \omega_{\vec q}$ the energy
of the phonon at momentum $\vec q$, and $\hat{q}=\vec{q}/|\vec{q}|$. 
In what follows, we consider the majorana-phonon coupling at the
\em{top} and \em{bottom} edges only. The Fourier transforms of the Majorana field $\eta(x)$ and of the disorder field along those edges $\zeta(x)$ are
\begin{align}
\label{eq:fouriertransformeta}
\eta(x)=\sqrt{\frac{2}{L_x}}\sum_{k}\eta_k e^{i k x},\quad
\zeta(x)=\frac{1}{\sqrt{L_x}}\sum_{k}\zeta_k e^{i k x},
\end{align}
where $L_x$ is the length of the system along the $x$ direction, such
that $\{\eta(x),\eta(x')\}=2\delta(x-x')$,
$\{\eta_k,\eta_{k'}\}=\delta_{k,-k'}$, $\la \zeta(x)\zeta(x')
\ra_{dis}=\zeta^2\,\delta(x-x')$, $\la \zeta_p\zeta_{-p'}\ra_{dis}=\la
\zeta_p\zeta_{p'}^*\ra_{dis} = \zeta^2 \delta_{p,p'}$, where
$\langle\cdots\rangle_{dis}$ denotes a disorder average. $\mc{H}_0$
and the phonon-Majorana interaction Hamiltonian $\mc{H}_{int}$ [from Eq.~(15)] in momentum space are
\begin{align}
\mc{H}_0&=\frac{1}{2}\sum_{k} v_f k\,
          \eta_{-k}\eta_{k}+\sum_{\vec{q}}\hbar\omega_{\vec{q}}\cd_{\vec
          q}c_{\vec q}\\
\mc{H}_{int}&=i\frac{g
              v_f}{4v_{ph}}\sqrt{\frac{\hbar}{2\rho_0}}\sqrt{\frac{1}{VL_x}}\sum_{\vec
              q, k, k', p}\sqrt{\omega_{\vec q}}(k-k')K_{ij}\hat
              q_i\hat q_j\left(\zeta_{p}\cd_{\vec
              q}\eta_k\eta_{k'}e^{-i q_y
              y_0}\delta_{p+k+k'-q_x}-\zeta_{p}c_{\vec q}\eta_k\eta_{k'}e^{i q_y y_0}\delta_{p+k+k'+q_x}\right)
\end{align}
where $y_0=L_y/2$, $\omega_{\vec q}=v_{ph}\sqrt{q_x^2+q_y^2}$, with the
acoustic phonon velocity $v_{ph}=\sqrt{B/\rho_0}$. Note that the
relation $\eta_k=\eta_{-k}^{\dg}$ has not been explicitly applied, and
the summation over $k, k'$ runs over both negative and positive
momenta [but only the annihilation mode $\eta_k$ appears in
Eq.~\eqref{eq:fouriertransformeta}] \footnote{Using the Fourier
  transform definition
  $\eta(x)=\sqrt{\frac{2}{L_x}}\sum_{k\geq0}\left(\eta_ke^{ikx}+\eta_{k}^\dagger
  e^{-ikx}\right)$ where
  the sum runs only over positive momenta, we obtain the same results.
}.

The scattering matrix element $\mc{M}^{\pm}$ that
{\em creates}, resp.\ {\em annihilates}, a single phonon mode (i.e.\
which multiplies $\cd_{\vec q}\eta_k\eta_{k'}$, resp.\ $c_{\vec q}\eta_k\eta_{k'}$) can be expressed as:
\begin{align}\label{eq:MatrixElement}
\mc{M}^+ (\vec q;k,k')&=i\frac{g
                                              v_f}{4v_{ph}}\sqrt{\frac{\hbar}{2\rho_0}}\sqrt{\frac{1}{VL_x}}K_{ij}\hat
                                              q_i\hat
                                              q_j\sqrt{\omega_{\vec q}}(k-k')\zeta_{q_x-k-k'}e^{-i
                                              q_y y_0}\non\\
\mc{M}^- (\vec q;k,k')&=-i\frac{g
                                            v_f}{4v_{ph}}\sqrt{\frac{\hbar}{2\rho_0}}\sqrt{\frac{1}{VL_x}}K_{ij}\hat
                                            q_i\hat
                                            q_j\sqrt{\omega_{\vec q}}(k-k')\zeta_{-q_x-k-k'}e^{i
                                            q_y y_0}.
\end{align}
To obtain the collision rate of a phonon mode at momentum $\vec q$ due to
the scattering with Majorana fermions, i.e.\ $\Big(\frac{\partial g
  (\vec q)}{\partial t}\Big)_{coll}$, we approximate the distribution
functions for the phonons and edge Majorana fermions as the thermal
distribution of bosons and fermions respectively, with different local
temperatures ${T}+\tilde T_{ph}$, ${T}+\tilde T_f$. This approximation should
presumably be valid at leading linear order in $\tilde T_{ph}-\tilde T_f$ since the
deviation from the thermal distribution should only contribute at
higher orders in $\tilde T_{ph}-\tilde T_f$. We argue that the system
reaches local thermalization due to phonon-phonon and phonon-Majorana
scattering. Moreover, given the absence of particle number
conservation for either the phonons or Majoranas, the chemical potentials $\mu_{ph,f}=0$ even away from equilibrium. From Fermi's golden rule, we have
\begin{align}\label{eq:CollisionRate}
\Big(\frac{\partial g (\vec q)}{\partial
  t}\Big)_{coll}&=\frac{2\pi}{\hbar}\times2\sum_{k,k'}\{\la|\mc{M}^+(\vec
                  q,k,k')|^2\ra_{dis}
                  (1+g_\omega)f_{\epsilon}f_{\epsilon'}\delta(-\omega_{\vec
                  q}+\epsilon_k+\epsilon_{k'})\non\\
&\quad\quad\quad\quad-\la|\mc{M}^-(\vec{q},k,k')|^2\ra_{dis}
  \,g_\omega f_{\epsilon} f_{\epsilon'}\delta(\omega_{\vec q}+\epsilon_k+\epsilon_{k'})\}\non\\
&=\frac{2\pi}{\hbar}\times2\sum_{k,k'}\{\la|\mc{M}^+(\vec q,k,k')|^2\ra_{dis}
  (1+g_\omega)f_{\epsilon}f_{\epsilon'}\delta(-\omega_{\vec q}+\epsilon_k+\epsilon_{k'})\non\\
&\quad\quad\quad\quad-\la|\mc{M}^-(\vec q,-k,-k')|^2\ra_{dis}
  \,g_\omega f_{-\epsilon} f_{-\epsilon'}\delta(\omega_{\vec q}-\epsilon_k-\epsilon_{k'})\}\non\\
&=\frac{2\pi}{\hbar}\times2\sum_{k,k'}\{\la|\mc{M}^+(\vec
  q,k,k')|^2\ra_{dis}\mathbb{P}(\omega_{\vec
  q},\epsilon_k,\epsilon_{k'})\delta(-\omega_{\vec
  q}+\epsilon_k+\epsilon_{k'}),
\end{align}
where
\begin{align}
  \label{eq:2}
\mathbb{P}(\omega_{\vec q},\epsilon_k,\epsilon_{k'})&=(1+g_\omega)f_{\epsilon}f_{\epsilon'}-g_\omega(1-f_{\epsilon})(1-f_{\epsilon'}),
\end{align}
with $g_\omega=1/(e^{\beta \omega}-1)$ and $f_\epsilon=1/(e^{\beta \epsilon}+1)=1-f_{-\epsilon}$. The factor of $2$ in front of the summation in Eq.~\eqref{eq:CollisionRate} comes from the two ways of creating and annihilating any given Majorana pair. The notation $\epsilon=\epsilon_k,\,\epsilon'=\epsilon_{k'},\,\omega=\omega_{\vec q}$ is used so long as there is no
ambiguity ($\epsilon_k$ is the Majorana fermion dispersion). The reality of the Majorana mode $\eta(x)$ requires that $\eta_k=\eta_{-k}^{\dg}$, thus $\epsilon_k=-\epsilon_{-k}$.
From the first to the second line, we take $k\rightarrow
-k,\,k'\rightarrow -k'$, i.e. $\epsilon\rightarrow - \epsilon$, which
is valid because the summation over $k,k'$ runs over both positive and negative
values in our convention. From the second to the last line in
Eq.~\eqref{eq:CollisionRate}, we used the fact that
$|\mc{M}^+(\vec q,k,k')|^2=|\mc{M}^-(\vec q,-k,-k')|^2$. The total rate of
energy change of the phonon subsystem through the collision with edge Majorana fermions is:
\begin{align}\label{eq:EnergyRate}
\Big(\frac{\partial \mc{E}}{\partial
  t}\Big)_{ph\rightarrow f}&=-\Big(\frac{\partial \mc{E}_{ph}}{\partial
  t}\Big)=-\sum_{q_x,q_y}\omega_{\vec q}\Big(\frac{\partial g (\vec q)}{\partial t}\Big)_{coll},
\end{align}
and the heat current from the phonons to the edge Majorana fermions is $$j_{ex}=\frac{1}{L_x}\Big(\frac{\partial \mc{E}}{\partial t}\Big)_{ph\rightarrow f}.$$

To obtain $\Big(\frac{\partial \mc{E}}{\partial t}\Big)_{ph\rightarrow f}$ at
leading order in the temperature difference $\tilde T_{ph}-\tilde T_f$
(i.e., leading order in $\delta\beta=\beta_f-\beta_{ph}$), we Taylor
expand $\mathbb{P}(\omega_{\vec q},\epsilon_k,\epsilon_{k'})$ to $\delta \beta$ order:
\begin{align}\label{eq:Distribution}
\mathbb{P}(\omega_{\vec q},\epsilon_k,\epsilon_{k'})=-\frac{\delta \beta\,  \omega }{2(\sinh \beta  \omega -\sinh \beta  (\epsilon -\omega )+\sinh \beta  \epsilon) }
\end{align}
where we made use of the energy conservation relation
$\epsilon'=\omega-\epsilon$. For simplicity, we consider isotropic
lattice distortions, and take $K_{ij}=\delta_{ij}$ in
Eq.~\eqref{eq:MatrixElement}, so that $\sum_{i,j}K_{ij}\hat q_i\hat
q_j=1$. Using
Eqs.~(\ref{eq:MatrixElement},\ref{eq:CollisionRate},\ref{eq:Distribution}),
we find:
\begin{align}\label{eq:EnergyRate}
\Big(\frac{\partial \mc{E}}{\partial
  t}\Big)_{ph\rightarrow f}&=-\frac{2\pi}{\hbar}\frac{\tilde\Gamma^2}{VL_x}\sum_{q_x,q_y,k,k'}2\,\omega_{\vec
           q}\Big(\sqrt{\omega_{\vec
           q}}(\epsilon_k-\epsilon_{k'})\Big)^2\la\zeta_{q_x-k-k'}\zeta_{q_x-k-k'}^*\ra_{dis}\mathbb{P}(\omega_{\vec
           q},\epsilon_k,\epsilon_{k'})\times\delta(-\omega_{\vec q}+\epsilon_k+\epsilon_{k'})\non\\
&=\frac{2\pi}{\hbar}\delta \beta\frac{\tilde\Gamma^2\zeta^2}{VL_x}\frac{L_x}{2\pi v_f}\sum_{q_x,q_y,k} \frac{\omega^3(2\epsilon-\omega)^2 }{\sinh \beta  \omega -\sinh \beta  (\epsilon -\omega )+\sinh \beta  \epsilon }
\end{align}
where $\tilde
\Gamma=\frac{g}{4v_{ph}}\sqrt{\frac{\hbar}{2\rho_0}}$. From the first
to the second line, the summation over $k'$ was performed with the
energy conservation $\delta(-\omega_{\vec
  q}+\epsilon_k+\epsilon_{k'})$, contributing a factor
$\frac{L_x}{2\pi v_f}$. The sum $\sum_{q_x,q_y,k}$ becomes, in the continuous limit,
\begin{align}
\sum_{q_x,q_y,k}\rightarrow \frac{L_x^2L_y}{(2\pi)^3}\int\diff
  q_x\diff q_y\diff k=\frac{ V L_x}{(2\pi)^3v_{ph}^2 v_f}\int
  \omega\diff \omega\diff \theta_{\vec q}\diff \epsilon.
\end{align}
Eq.~\eqref{eq:EnergyRate} then becomes:
\begin{align}\label{eq:EnergyRate1}
\Big(\frac{\partial \mc{E}}{\partial
  t}\Big)_{ph\rightarrow f}=\delta
  \beta\frac{2\pi}{\hbar}\frac{\tilde\Gamma^2\zeta^2}{VL_x}\frac{L_x}{2\pi
  v_f}\frac{ V L_x}{(2\pi)^3v_{ph}^2 v_f}\int_0^{\Lambda} \diff
  \omega\int_{0}^{2\pi}\diff \theta_{\vec q}\int_{-\Lambda'}^{\Lambda'}\diff \epsilon \, \frac{ \omega^4(2\epsilon-\omega)^2 }{\sinh \beta  \omega -\sinh \beta  (\epsilon -\omega )+\sinh \beta  \epsilon },
\end{align}
where $\Lambda$ and $\Lambda'$ are the phonon and Majorana fermion
energy cutoffs, respectively. From Eq.~\eqref{eq:EnergyRate1}, by power counting energies $\omega,\epsilon$, we can already see that the result takes the form of
\begin{align}\label{eq:current}
j_{ex}=\frac{1}{L_x}\Big(\frac{\partial \mc{E}}{\partial
  t}\Big)_{ph\rightarrow f}=\delta
  \beta\frac{1}{\hbar}\frac{\tilde\Gamma^2\zeta^2}{(2\pi)^3v_{ph}^2
  v_f^2}\left(\frac{1}{\beta}\right)^8 f(v_{ph}/v_f)=\delta
  \beta\frac{g^2\zeta^2}{32(2\pi)^3}\frac{1}{v_{ph}^4
  v_f^2\rho_0}\left(\frac{1}{\beta}\right)^8 f(v_{ph}/v_f)\sim T^6 (\tilde T_{ph}-\tilde T_f),
\end{align}
where $\delta \beta/\beta^2=\tilde T_{ph}-\tilde T_{f}$. Importantly, from $j_{ex}\sim T^6 (\tilde T_{ph}-\tilde T_f)$, we can extract the exponent $\alpha$ of the thermal coupling $\lambda(T)$ defined in the main text, i.e., $\alpha=6$. In passing, we note that without disorder, two-phonon processes are necessary to satisfy kinematic constraints, and one obtains an even larger $\alpha=8$. The function $f(x)$ in general depends on the ratio of $v_{ph}/v_f$, and is a dimensionless constant in this model:
\begin{align}\label{eq:f}
f=&\int_0^{\infty} \diff {\sf u}\int_{0}^{2\pi}\diff \theta_{\vec
    q}\int_{-\infty}^{\infty}\diff {\sf v}\, (2{\sf v}-{\sf u})^2\times \frac{
    {\sf u}^4 }{\sinh   {\sf u} -\sinh   ({\sf v} -{\sf u} )+\sinh
    {\sf v} }\approx 4.2\times 10^4,
\end{align}
where ${\sf u}=\beta\omega$, ${\sf v}=\beta \epsilon$, and at low enough
temperature, we take the integration range such that
$\Lambda\beta,\Lambda'\beta\rightarrow \infty$. The calculation of $f$
is outlined below, To simplify the integral, we note that the
Boltzmann distribution term
$\mathbb{P}(\omega,\epsilon,\omega-\epsilon)$ is symmetric around
$\epsilon=\omega/2$, so we makes the change of variables ${\sf v}-{\sf
  u}/2\rightarrow\tilde {\sf v}$. The integral measure and range remain unchanged, and the integrand in Eq.~\eqref{eq:f} becomes
\begin{align} \label{eq:f2}
f=&\int_0^{\infty} \diff {\sf u}\int_{0}^{2\pi}\diff
    \theta_{\vec q}\int_{-\infty}^{\infty}\diff \tilde {\sf v}\,\,4\tilde
    {\sf v}^2\frac{{\sf u}^4 }{2 \sinh\frac{{\sf u} }{2}\left(\cosh \frac{{\sf u}
    }{2}+\cosh \tilde {\sf v}\right)}.
\end{align}
The integral over $\tilde {\sf v}$ can be straightforwardly performed using
the result:
\begin{align}
\int_{-\infty}^{\infty} \diff \tilde {\sf v}\frac{\tilde {\sf
  v}^2}{\cosh \tilde {\sf v}+\cosh \ell }&=\frac{2\ell(\pi^2+\ell^2) }{3\sinh \ell}.
\end{align}
Integrating over ${\sf u}$, we obtain finally:
\begin{align}
f=2\pi\times160\times \big(2\pi^2 \zeta (5)+21 \zeta (7)\big)\approx4.2\times 10^4,
\end{align}
where $m\mapsto\zeta (m)$ is the Riemann zeta function.

\subsection{A more microscopic derivation of the spin-lattice-disorder
  vertex}
\label{sec:EffectiveVertex}

In this section, we consider a more generic form of phonon-Majorana coupling ($\mc{H}_{f-ph}$) and disorder-Majorana coupling ($\mc{H}_{f-dis}$) at the edge at $y=y_0=L_y/2$ as:
\begin{align}\label{eq:coupling0}
\mc{H}_{f-ph}&=\frac{ig v_f }{4}\int \diff x\diff y K_{ij}\partial_iu_j(x,y)\eta(x)\partial_x\eta(x)\delta (y-y_0),\non\\
\mc{H}_{f-dis}&=\frac{-i v_f }{4}\int \diff x \zeta(x)\eta(x)\partial_x\eta(x).
\end{align}
Our goal is to justify the disorder vertex taken in
Eq.~(15) by incorporating the velocity disorder
$\mc{H}_{f-dis}$ into the vertex disorder $\sim \mc{H}_{int}$ by
properly rescaling the Majorana field in a way that preserves the
anticommutation relation. The quadratic Hamiltonian of the Majorana
field is now
\begin{align}
\mc{H}_{quad}&=\frac{-i v_f}{4} \int \diff x\, (1+\zeta(x))\eta(x)\partial_x\eta(x).
\end{align}
We define the following canonical transformation of the Majorana field:
\begin{align}
\eta(x)=\left|\frac{\partial \tilde x (x)}{\partial x}\right|^{1/2}\tilde \eta(\tilde x),
\end{align}
for any function $\tilde x(x)$ that is invertible, i.e.\ $\{\tilde \eta(\tilde x),\tilde \eta(\tilde x')\}=2\delta(\tilde x-\tilde x')$. Considering weak disorder, a natural choice is $\partial \tilde x (x)/\partial x>0$ and $\tilde x(x)\rightarrow \pm \infty $ as $x\rightarrow \pm \infty$. In terms of the new coordinate,
\begin{align}
\eta\partial_x\eta=\left|\frac{\partial \tilde x (x)}{\partial x}\right|^2\tilde\eta\partial_{\tilde x}\tilde\eta,\quad \int\diff x=\int\diff\tilde x \left|\frac{\partial \tilde x (x)}{\partial x}\right|^{-1}.
\end{align}
Thus $\mc{H}_{quad}$ becomes
\begin{align}
\mc{H}_{quad}&=\frac{-i v_f}{4} \int \diff \tilde x\, (1+\zeta(x))\left|\frac{\partial \tilde x (x)}{\partial x}\right|\tilde \eta(\tilde x)\partial_{\tilde x}\tilde\eta(\tilde x).
\end{align}
To remove the disorder vertex in $\mc{H}_{quad}$, we require
$\left|\frac{\partial \tilde x (x)}{\partial
    x}\right|=\frac{1}{1+\zeta(x)}$. Instead, $\zeta(x)$ appears in
the Majorana-phonon vertex as follows:
\begin{align}\label{eq:vertex1}
\mc{H}_{f-ph}=\frac{ig v_f}{4} \int \diff \tilde x\diff y \,\frac{K_{ij}\partial_iu_j(x,y)}{1+\zeta(x)}\tilde \eta(\tilde x)\partial_{\tilde x}\tilde\eta(\tilde x)\delta (y-y_0).
\end{align}
To express $\frac{K_{ij}\partial_iu_j(x,y)}{1+\zeta(x)}$ in terms of
$\tilde x$ to leading order in $\zeta$ (weak disorder limit),
\begin{align}
\tilde x (x)=\int^x\diff x'\frac{1}{1+\zeta(x')}= \int^x\diff x'(1-\zeta(x'))+\mc{O}(\zeta^2)=x-\int^x\diff x' \zeta(x')+\mc{O}(\zeta^2).
\end{align}
The inverse transformation is
\begin{align}
x(\tilde x)=\tilde x+\int^x\diff x' \zeta(x')=\tilde x+\int^{\tilde x+\int^x\diff x'' \zeta(x'')}\diff x' \zeta(x')=\tilde x+\int^{\tilde x}\diff x' \zeta(x')+\mc{O}(\zeta^2).
\end{align}
For $K_{ij}\partial_iu_j(x,y)$ in terms of $(\tilde x, \tilde y=y)$, with a careful analysis for $i=x$ and $i=y$ separately at the leading order in $\zeta$:
\begin{align}
K_{xj}\partial_xu_j(x,y)&=K_{xj}\frac{\partial \tilde x}{\partial x}\partial_{\tilde x}u_j(x(\tilde x),y)=K_{xj}(1-\zeta(\tilde x))\partial_{\tilde x}u_j(\tilde x+\int^{\tilde x}\diff x' \zeta(x'))\non\\
&=K_{xj}(1-\zeta(\tilde x))\partial_{\tilde x}[u_j(\tilde x)+\partial_{\tilde x}u_j(\tilde x)\int^{\tilde x}\diff x' \zeta(x')]\non\\
&=K_{xj}(1-\zeta(\tilde x))[\partial_{\tilde x}u_j(\tilde x)+\zeta(\tilde x)\partial_{\tilde x}u_j(\tilde x)+\partial^2_{\tilde x}u_j(\tilde x)\int^{\tilde x}\diff x' \zeta(x')]\non\\
&=K_{xj}(\partial_{\tilde x}u_j(\tilde x)+\partial^2_{\tilde x}u_j(\tilde x)\int^{\tilde x}\diff x' \zeta(x')),\non\\
K_{yj}\partial_y u_j(x,y)
&=K_{yj}\partial_{y}u_j(\tilde x+\int^{\tilde x}\diff x' \zeta(x'))\non\\
&=K_{yj}(\partial_{y}u_j(\tilde x)+\partial_{\tilde x}\partial_y u_j(\tilde x)\int^{\tilde x}\diff x' \zeta(x'))\non\\
&=K_{yj}(\partial_{\tilde y}u_j(\tilde x)+\partial_{\tilde x}\partial_{\tilde y} u_j(\tilde x)\int^{\tilde x}\diff x' \zeta(x')),
\end{align}
we have $K_{ij}\partial_iu_j(x,y)=K_{ij}(\partial_{\tilde i}u_j(\tilde x)+\partial_{\tilde x}\partial_{\tilde i} u_j(\tilde x)\int^{\tilde x}\diff x' \zeta(x'))+\mc{O}(\zeta^2)$. The $y$ coordinate in the expression of $u_j(x,y)$ is omitted for brevity so long as there is no ambiguity.
To leading linear order in $\zeta$, Eq.~\eqref{eq:vertex1} is
\begin{align}\label{eq-b28}
\mc{H}_{f-ph}&=\frac{ig v_f}{4} \int \diff \tilde x\diff y \,K_{ij}\partial_iu_j(x,y)(1-\zeta(x))\tilde \eta(\tilde x)\partial_{\tilde x}\tilde\eta(\tilde x)\delta (y-y_0)\non\\
&=\frac{ig v_f}{4} \int \diff \tilde x\diff y \,K_{ij}(\partial_{\tilde i}u_j(\tilde x)+\partial_{\tilde x}\partial_{\tilde i} u_j(\tilde x)\int^{\tilde x}\diff x' \zeta(x'))(1-\zeta(\tilde x))\tilde \eta(\tilde x)\partial_{\tilde x}\tilde\eta(\tilde x)\delta (y-y_0)\non\\
&= \frac{ig v_f}{4}\int \diff \tilde x\diff y \,\left(K_{ij}\partial_{\tilde i}u_j(\tilde x)-K_{ij}\partial_{\tilde i}u_j(\tilde x)\zeta(\tilde x)+\partial_{\tilde x}(K_{ij}\partial_{\tilde i}u_j(\tilde x))\int^{\tilde x}\diff x' \zeta(x')\right)\tilde \eta(\tilde x)\partial_{\tilde x}\tilde\eta(\tilde x)\delta (y-y_0).
\end{align}
Note that the second term in the last line is exactly the disorder vertex in Eq.~(15), while the last term is subleading in the limit $v_{ph}/v_f\gg 1$ despite its integral form. To show that, we first specify the boundary condition for the coordinate transformation as $\zeta(\pm \infty)=0$, and the Fourier transform of the disorder integral in the last term can be well defined
\begin{align}
\int^{\tilde x}\diff x' \zeta(x')=\int_{-\infty}^{\tilde x}\diff x' \zeta(x')=\frac{1}{\sqrt{L_x}}\sum_p\frac{\zeta_p \,e^{i p \tilde x}}{i p}.
\end{align}
The Fourier transform of the last term becomes
\begin{align}
&\frac{ig v_f}{4}\int \diff \tilde x\diff y \,\left(\partial_{\tilde x}(K_{ij}\partial_{\tilde i}u_j(\tilde x))\int^{\tilde x}\diff x' \zeta(x')\right)\tilde \eta(\tilde x)\partial_{\tilde x}\tilde\eta(\tilde x)\delta (y-y_0)\non\\
=&-i\frac{g v_f}{4v_{ph}}\sqrt{\frac{\hbar}{2\rho_0}}\sqrt{\frac{1}{VL_x}}\sum_{\vec q,k,k'}K_{ij}\hat q_i\hat q_j\sqrt{\omega_{\vec q}}\,(k-k')\frac{q_x}{q_x-k-k'}\zeta_{q_x-k-k'}e^{-iq_y y_0}\cd_{\vec q}\eta_k\eta_{k'}+h.c.,
\end{align}
which only differs from that of Eq.~\ref{eq:MatrixElement} by a factor
$-\frac{q_x}{q_x-k-k'}$. Since $(k+k')=(v_{ph}/v_f)\,
\sqrt{q_x^2+q_y^2}=(v_{ph}/v_f)\, q_x/\cos\theta_{\vec q}$ from energy
conservation, the factor $-\frac{q_x}{q_x-k-k'}=\frac{\cos\theta_{\vec
    q}v_f/v_{ph}}{1-\cos\theta_{\vec q}v_f/v_{ph}}\sim v_f/v_{ph}\ll
1$ for realistic values. This implies that the last term
in Eq.~\eqref{eq-b28} with the disorder integral is subleading compared to the
disorder vertex discussed in the main text \footnote{The same conclusion can be obtained by calculating the matrix element diagrammatically [see Fig.~\ref{fig-3}(b)] from the couplings of Eq.~\eqref{eq:coupling0}. }.

\end{document}